\documentclass[12pt]{article}
 \usepackage{graphicx}
 \usepackage[cp1251]{inputenc}
 \usepackage{rotating}
\begin{document}
 \noindent {\footnotesize\it Astronomy Letters, 2004, Vol. 30, No. 12, pp. 848--853.}
 \newcommand{\dif}{\textrm{d}}

 \noindent
 \begin{tabular}{llllllllllllllllllllllllllllllllllllllllllllll}
 & & & & & & & & & & & & & & & & & & & & & & & & & & & & & & & & & & & & & \\\hline\hline
 \end{tabular}

 \vskip 0.5cm
 \centerline {\bf ASTROMETRIC CONTROL  OF THE INERTIALITY }
 \centerline {\bf  OF THE HIPPARCOS CATALOG}
 \bigskip
 \centerline {\bf V.V.~Bobylev}
  \bigskip
  \centerline{\small\it Pulkovo Astronomical Observatory, St. Petersburg,  Russia}
  \bigskip
  \bigskip
{\bf Abstract}—Based on the most complete list of the results of
an individual comparison of the proper motions for stars of
various programs common to the Hipparcos catalog, each of which is
an independent realization of the inertial reference frame with
regard to stellar proper motions, we redetermined the vector
${\hbox {\boldmath $\omega$}}$ of residual rotation of the ICRS
system relative to the extragalactic reference frame. The
equatorial components of this vector were found to be the
following:
 $\omega_x = +0.04\pm0.15$ mas yr$^{-1}$,
 $\omega_y = +0.18\pm0.12$ mas yr$^{-1}$, and
 $\omega_z = -0.35\pm0.09$ mas yr$^{-1}$.


\bigskip
\bigskip
\leftline {\hskip6mm INTRODUCTION}
\bigskip
A kinematic analysis performed previously (Bobylev 2004) revealed
appreciable residual rotation of the stars of the Hipparcos
catalog (ESA 1997) with respect to the inertial reference frame
around the Galactic y axis with an angular velocity of
 $-0.36\pm0.09$ mas yr$^{-1}$ (milliarcseconds per year). A shortcoming of
the method used is the absence of a rigorous criterion for
separating the actual rotation of stars close to the Sun from the
sought systematic rotation. The method of analyzing the
proper-motion differences for stars of various programs common to
the Hipparcos catalog, each of which is an independent realization
of the inertial reference frame with regard to stellar proper
motions, is free from this shortcoming. We call this method
astrometric. Based on this method, Kovalevsky et al. (1997) found
that all three components of the vector
 ${\hbox {\boldmath $\omega$}}$
 of residual
rotation of the Hipparcos catalog with respect to the
extragalactic reference frame have no significant deviations from
zero, with the error of this vector along the three axes being
$\pm0.25$ mas yr$^{-1}$. The difficulties in applying this method
stem from the fact that there are few VLBI observable radio stars,
while photographic catalogs are not free from the magnitude
dependence of the stellar proper motions (the magnitude equation).
The goal of this study is to redetermine the vector
 ${\hbox {\boldmath $\omega$}}$
using the most complete list of the independent results from an
individual comparison of the stellar proper motions.

\bigskip
\leftline {\hskip6mm CHARACTERISTIC OF INDIVIDUAL SOLUTIONS}
\bigskip
Based on the NPM photographic program of the Lick observatory
(Klemola et al. 1987), two catalogs of absolute proper motions for
northern-sky stars have been published: the NPM1 catalog of stars
in 899 areas outside the Milky Way zone (Klemola et al. 1994) and
the NPM2 catalog of stars in 347 areas of the Milky Way zone
(Hanson et al. 2003). The areas of the NPM1 and NPM2 catalogs do
not overlap. The results of a comparison of the proper motions for
NPM1 and Hipparcos stars were described by Platais et al. (1998b)
and designated as NPM (Yale) by Kovalevsky et al. (1997). An
independent analysis of the proper-motion differences between the
NPM1 and Hipparcos catalogs was also performed in Heidelberg;
Kovalevsky et al. (1997) designated the results of this analysis
as NPM (Heidelberg). All of the above authors agree that the
component $\omega_z$ is difficult to determine using the NPM1
catalog, because there is a magnitude equation in the NPM1 proper
motions that causes the NPM1–Hipparcos differences to be shifted
by $\approx6$ mas yr$^{-1}$ for the brightest ($\approx8^m$) and
faintest ($\approx12^m$) stars. Kovalevsky et al. (1997) did not
include the component $\omega_z$ determined from NPM1 stars in
their final solution.

Zhu (2003) compared the proper motions of NPM2 and Hipparcos
stars. This author found no appreciable magnitude dependence of
the NPM2--Hipparcos differences.

Based on the SPM program of photographic observations of
southern-sky stars (Platais et al. 1995),
 Platais et al. (1998a) published the SPM2
catalog of absolute proper motions of stars in 156 southern sky
areas. The results of a comparison of the absolute proper motions
of SPM stars in 63 areas with the Hipparcos catalog were described
by Platais et al. (1998a). We designate these SPM stars as SPM1.
Zhu (2001) compared the proper motions of SPM2 and Hipparcos
stars.

Based on the combined photographic catalog GPM (Rybka and Yatsenko
1997a), Rybka and Yatsenko (1997b) published a list of proper
motions for bright stars common to the Hipparcos catalog
designated as GPM1. Kislyuk et al. (1997) compared the proper
motions of GPM1 and Hipparcos stars. These authors concluded that
there is a magnitude equation in the GPM1 catalog, and the
parameters $\omega_x$, $\omega_y$, and $\omega_z$ determined only
from faint comparison stars must be used.

Bobylev et al. (2004) determined the parameters $\omega_x$,
$\omega_y$, and $\omega_z$ by comparing the Pulkovo photographic
catalog PUL2 and Hipparcos and found no appreciable magnitude
equation in the proper motions of PUL2 stars. Each of the GPM and
PUL2 catalogs is an independent realization of the photographic
plan by Deutch (1954); the centers of the areas coincide and
correspond to the list of Deutch (1955).

The results of the well-known independent programs were presented
by Kovalevsky et al. (1997). In addition, the results of the
Potsdam program were presented by Hirte et al. (1996), and the
Bonn program by Geffert et al. (1997) and Tucholke et al. (1997);
the Earth Rotation Parameters (EOP) were analyzed by Vondrak et
al. (1997); radio stars were analyzed by Lestrade et al. (1995),
and the Hubble Space Telescope (HST) observations were analyzed by
Kovalevsky et al. (1997).

{\footnotesize\begin{table}[t] \caption[]
{\small\baselineskip=1.0ex\protect Components of the vector {\hbox
{\boldmath $\omega$}} --- $\omega_x,\omega_y,$ and $\omega_z$ (in
mas yr$^{-1}$) that we determined from the VLA--Hipparcos
differences using data from Boboltz et al. (2003). }
 \medskip
{\small\begin{center}\parbox{12truecm}{\offinterlineskip\halign
{#\vrule&#\hfil&\vrule#&\hfil#&\vrule#&\hfil#&#&\hfil#&\hfill#&
#\hfil&\hfil#&\hfil#&\vrule#\cr\noalign{\hrule}&&&&&&&&&&&&height6pt\cr
&~ Number of stars
&&\hfill$\sigma_\circ$\hfill&&&&&&\hfill$\omega_x$\hfill&\hfill$\omega_y$\hfill&\hfill$\omega_z$\hfill&\cr&&&&&&&&&&&&height6pt\cr\noalign{\hrule}
&&&&&&&&&&&&height6pt\cr &~ 15 (all)     &&~ $\pm2.06$
~&&&&&&~$-0.97\pm0.62$ ~& $-1.59\pm0.73$ ~&$+0.18\pm0.63$
~&\cr&&&&&&&&&&&&height6pt\cr &~ 12 (*)       &&~ $\pm1.63$
~&&&&&&~$-0.42\pm0.56$ ~& $-0.51\pm0.64$ ~&$+0.20\pm0.57$
~&\cr&&&&&&&&&&&&height6pt\cr
\noalign{\hrule}}}%
\end{center}}
{\footnotesize
 (*) The radio stars HD 50896 N, KQ Pup, and RS CVn were rejected. }
\end{table}}

The main difference between our approach and the approach of
Kovalevsky et al. (1997) is that we included the following data:

(1) The results of a comparison of the proper motions for PUL2 and
Hipparcos stars (Bobylev et al. 2004);

(2) The results of a comparison of the proper motions for SPM2 and
Hipparcos stars (Zhu 2001);

(3) The results of a comparison of the proper motions for NPM2 and
Hipparcos stars (Zhu 2003);

(4) The results of a comparison of the VLA absolute proper motions
for radio stars and their Hipparcos proper motions.

Table 1 gives the components of the vector ${\hbox {\boldmath
$\omega$}}$ that we calculated using the absolute proper motions
of 15 radio stars from Boboltz et al. (2003). These authors
described in detail the project and the method of referencing the
observations of radio stars to quasars and compared these stars
with the Hipparcos catalog, but did not determine the components
of the vector ${\hbox {\boldmath $\omega$}}$. We used the
equations in the form that was proposed and used by Lindegren and
Kovalevsky (1995):
$$
\displaylines{\hfill
 \Delta\mu_\alpha\cos\delta= \omega_x\cos\alpha\sin\delta
 +\omega_y\sin\alpha\sin\delta- \omega_z\cos\delta,
 \hfill\llap(1)\cr\hfill
 \Delta\mu_\delta=-\omega_x\sin\alpha+ \omega_y\cos\alpha,
 \hfill\llap(2)
 }
$$
where the Hipparcos catalog differences are on the left-hand sides
of the equations. Here, $\sigma_\circ$ is the error per unit
weight in the solution of Eqs.~(1) and (2). The VLA (Boboltz et
al. 2003) and VLBI (Lestrade et al. 1995, 1999) absolute proper
motions are independent.

Our approach also differs from that of Kovalevsky et al. (1997) in
that we used the results of a comparison of the NPM1 and Hipparcos
stellar proper motions performed by the Heidelberg team, which
were obtained from the data for 2616 stars, to determine
$\omega_z$. The random errors in $\omega_x$, $\omega_y$, and
$\omega_z$ for the Heidelberg solution are given only for the
sample of 1135 stars ($e_{\omega_x} = 0.25$ mas yr$^{-1}$ and
$e_{\omega_y} = e_{\omega_z} = 0.2$ mas yr$^{-1}$).
The corresponding errors for the sample of 2616 stars are smaller
by a factor of $\approx\sqrt{2616/1135}$. Since there are problems
with the NPM1 catalog, we assumed the random errors to be 0.25 mas
yr$^{-1}$ for $\omega_x$ and 0.20 mas yr$^{-1}$ for $\omega_y$ and
$\omega_z$ in order not to overestimate this solution. Our choice
of this solution was dictated by the fact that it was obtained in
the magnitude range $10.5^m-11.5^m$. As can be seen in Fig. 1c
from Platais et al. (1998b), the Hipparcos--NPM1 proper-motion
differences have a horizontal pattern near zero precisely in this
magnitude range. In our opinion, the NPM1 proper motions are least
affected by the magnitude equation in this magnitude range. In
addition, we used only one of the available (not independent)
results of a comparison of the SPM1 and Hipparcos catalogs, the
YVA solution (Kovalevsky et al. 1997; Platais et al. 1998b).

\bigskip
 \leftline {\hskip6mm DETERMINING THE VECTOR ${\hbox {\boldmath $\omega$}}$}
 \bigskip
We assigned a weight inversely proportional to the square of the
error e in the corresponding quantities $\omega_x$, $\omega_y$,
and $\omega_z$ to each comparison catalog, which was calculated
using the formula
$$
\displaylines{\hfill P_i={e_{kiev}}^2/{{e_i}^2}, \quad
i=1,...,12,\hfill\llap(3) }
$$
where $i$ is the number of individual sources, and $e_{kiev}$, the
random error of the GPM1 (Kiev) program. Table 2 contains the data
used here. The second column of the table gives the weights
calculated using formula (3). Not all of the authors use equations
in the form (1) and (2). In such cases, we reduced the signs of
the quoted quantities (Zhu 2001, 2003) to the same form.

{\footnotesize\begin{table}[t] \caption[]
{\small\baselineskip=1.0ex\protect (Equatorial) components of the
vector of residual rotation of the Hipparcos catalog with respect
to extragalactic objects --- $\omega_x$, $\omega_y$, and
$\omega_z$ (mas yr$^{-1}$).
 }
 \medskip
{\small\begin{center}\parbox{12truecm}{%
\offinterlineskip\halign
{#\vrule&#\hfil&\vrule#&\hfil#&\vrule#&\hfil#&\vrule#&\hfil#&\vrule#&
 #\hfil&\hfil#&\hfil#&\vrule#\cr\noalign{\hrule}&&&&&&&&&&&&height6pt\cr
&~ ~&&\hfill$P_x,P_y,P_z$\hfill&&~$N_\star$~&&~$N_{\hbox {\tiny
area}}$~&&\hfill$\omega_x$\hfill&\hfill$\omega_y$\hfill&\hfill$\omega_z$\hfill&\cr&&&&&&&&&&&&height6pt\cr\noalign{\hrule}&&&&&&&&&&&&height6pt\cr
&~NPM2       &&~16.00/4.62/8.16~&&~3519~&&~347~&&~ $-0.11\pm0.20$
~& $-0.19\pm0.20$ ~&$-0.75\pm0.28$ ~&\cr&&&&&&&&&&&&height4pt\cr
&~SPM2       &&~22.15/9.43/28.44~&&~9356~&&156~&&~ $+0.10\pm0.17$
~& $+0.48\pm0.14$ ~&$-0.17\pm0.15$ ~&\cr&&&&&&&&&&&&height4pt\cr
&~NPM1       &&~10.24/4.62/16.00~&&~2616~&&~899~&&~ $-0.76\pm0.25$
~& $+0.17\pm0.20$ ~&$-0.85\pm0.20$ ~&\cr&&&&&&&&&&&&height4pt\cr
&~SPM1       &&~44.44/5.71/130.61~&&~4067~&& 63~&&~
$+0.44\pm0.12$~& $+0.71\pm0.18$ ~&$-0.30\pm0.07$
~&\cr&&&&&&&&&&&&height4pt\cr &~PUL2       &&~2.90/1.28/3.63~&&~
1004~&& 147~&&~ $-0.98\pm0.47$ ~& $-0.03\pm0.38$ ~&$-1.66\pm0.42$
~&\cr&&&&&&&&&&&&height4pt\cr &~Kiev(GPM1) &&~1.0/1.0/1.0~&&~
415~&& 154~&&~  $-0.27\pm0.80$ ~& $+0.15\pm0.60$ ~&$-1.07\pm0.80$
~&\cr&&&&&&&&&&&&height4pt\cr &~Potsdam    &&~2.37/0.74/2.78~&&~
256~&& 24~&&~  $+0.22\pm0.52$ ~& $+0.43\pm0.50$ ~&$+0.13\pm0.48$
~&\cr&&&&&&&&&&&&height4pt\cr &~Bonn       &&~5.54/2.96/5.88~&&~
88~&& 13~&&~  $+0.16\pm0.34$ ~& $-0.32\pm0.25$ ~&$+0.17\pm0.33$
~&\cr&&&&&&&&&&&&height4pt\cr &~VLBI       &&~7.11/2.74/7.11~&&~
12~&&    &&~  $-0.16\pm0.30$ ~& $-0.17\pm0.26$ ~&$-0.33\pm0.30$
~&\cr&&&&&&&&&&&&height4pt\cr &~VLA+PT     &&~2.04/0.45/1.97~&&~
12~&&    &&~  $-0.42\pm0.56$ ~& $-0.51\pm0.64$ ~&$+0.20\pm0.57$
~&\cr&&&&&&&&&&&&height4pt\cr &~HST        &&~0.08/0.08/0.05~&&~
46~&&    &&~  $-1.60\pm2.87$ ~& $-1.92\pm1.54$ ~&$+2.26\pm3.42$
~&\cr&&&&&&&&&&&&height4pt\cr &~EOP        &&~8.16/2.36/---~~~&&~
~&&    &&~  $-0.93\pm0.28$ ~&  $-0.32\pm0.28$ ~&~~~   ---
~~~~~~~&\cr&&&&&&&&&&&&height6pt\cr\noalign{\hrule}&&&&&&&&&&&&height4pt\cr
 &~Mean~1 &&&&&&&&~$+0.04\pm0.15$ ~& $+0.18\pm0.12$~&$-0.35\pm0.09$~&\cr&&&&&&&&&&&&height4pt\cr
 &~Mean~2 &&&&&&&&~$-0.19\pm0.16$ ~& $+0.08\pm0.11$ ~&$-0.43\pm0.15$~&\cr&&&&&&&&&&&&height6pt\cr
\noalign{\hrule}}}%
\end{center}}
{\footnotesize Note: $N_\star$ is the number of common comparison
stars, and $N_{area}$ is the number of comparison areas common to
the photographic catalogs. Mean~2 was calculated without SPM1.}
\end{table}}

The lower part of Table~2 gives the weighted means of $\omega_x$,
$\omega_y$, and $\omega_z$. Mean 1 was calculated using all of the
available data in Table 2. The SPM1 and SPM2 catalogs are not
independent, but the methods of obtaining the solutions differ.
The SPM1 solution (Platais et al. 1998b) was obtained only from
the differences in $\mu_\alpha \cos \delta$ (e.g., only from Eq.
(1)). The SPM2 solution (Zhu 2001) was obtained by simultaneously
solving Eqs.~(1) and (2); the author pointed out that there is a
color equation in the SPM2 catalog, which was not eliminated, and
it is more pronounced in the $\mu_\delta$ differences. We
calculated Mean 2 without using the SPM1 solution in order to
analyze only independent sources. In this case, the SPM proper
motions also have the largest weight when calculating the
component $\omega_z$, as can be seen from the table. A comparison
of Means 1 and 2 reveals no significant differences between these
two solutions, with $\omega_z$ being an appreciable component. The
errors of the vector ${\hbox {\boldmath $\omega$}}$ along the
three axes
$e_{\omega}=\sqrt{{e_{\omega_x}}^2+{e_{\omega_y}}^2+{e_{\omega_z}}^2}$
are $\pm0.21$ and $\pm0.25$ mas yr$^{-1}$ for Mean~1 and 2
solutions, respectively. Our solutions for the components
$\omega_x$ and $\omega_y$ do not differ significantly from the
final solutions of Kovalevsky et al. (1997) based on both
Lindegren’s and Kovalevsky’s methods. At the same time, there is a
significant difference in the determination of $\omega_z$, which
we found to differ significantly from zero. The figure shows the
projections of the individual solutions for the vector ${\hbox
{\boldmath $\omega$}}$ onto the $xy, xz,$ and $yz$ planes based on
the data from Table 2. Also shown in the figure are the components
of our Mean~1 solution.

\bigskip
\leftline {\hskip6mm DISCUSSION}
\bigskip
The effects of the actual motions of stars in the differences
between the catalogs under consideration are ruled out. Therefore,
the causes of the rotation $\omega_z$ found are the following:

(1) Inaccurate realization of the ICRS or, in other words,
residual rotation of the Hipparcos catalog with respect to the
ICRF (Ma et al. 1998), and, in this case, the rotation found is of
a ``technical'' nature;

(2) Residual rotation of the ICRF itself with respect to the
extragalactic reference frame. The ICRF is based on ground-based
VLBI observations of extragalactic radio sources. It may be
assumed that, in this case, the rotation $\omega_z$ found is
precessional in nature and depends on the accuracy of the
constants used. Given the form of the proper-motion differences
and the signs on the right-hand sides of Eqs. (1)--(2), we can
write
$$
\displaylines{\hfill
 \omega_y=-\Delta p_1 \sin \varepsilon,\hfill\llap(4)\cr\hfill
 \omega_z= \Delta p_1 \cos \varepsilon-\Delta E,\hfill\llap(5)
 }
$$
where $\Delta p_1$ is the correction to the adopted constant of
lunisolar precession in longitude, $\Delta E$ is the sum of the
corrections to the rate of planetary precession and the motion of
the zero point of right ascensions, and $\varepsilon$ is the
inclination of the ecliptic to the equator. Using only Eq. (5) and
setting $\Delta E = 0$, we obtain from our Mean 1 solution
$$
 \displaylines{\hfill \Delta p_1=-0.38\pm0.10~{\hbox {mas yr$^{-1}$}}.\hfill\llap(6) \cr}
$$
This value agrees with the result of our previous work (Bobylev
2004), where we found $\Delta p_1 = -0.42\pm0.10$ mas yr$^{-1}$
from a kinematic analysis of the proper motions for distant
Hipparcos stars. In our opinion, a comparison of the residual
rotations around the ecliptic axis found by two independent
methods is most justifiable.

 \begin{figure}[p] {\begin{center}
 \includegraphics[width=80mm]{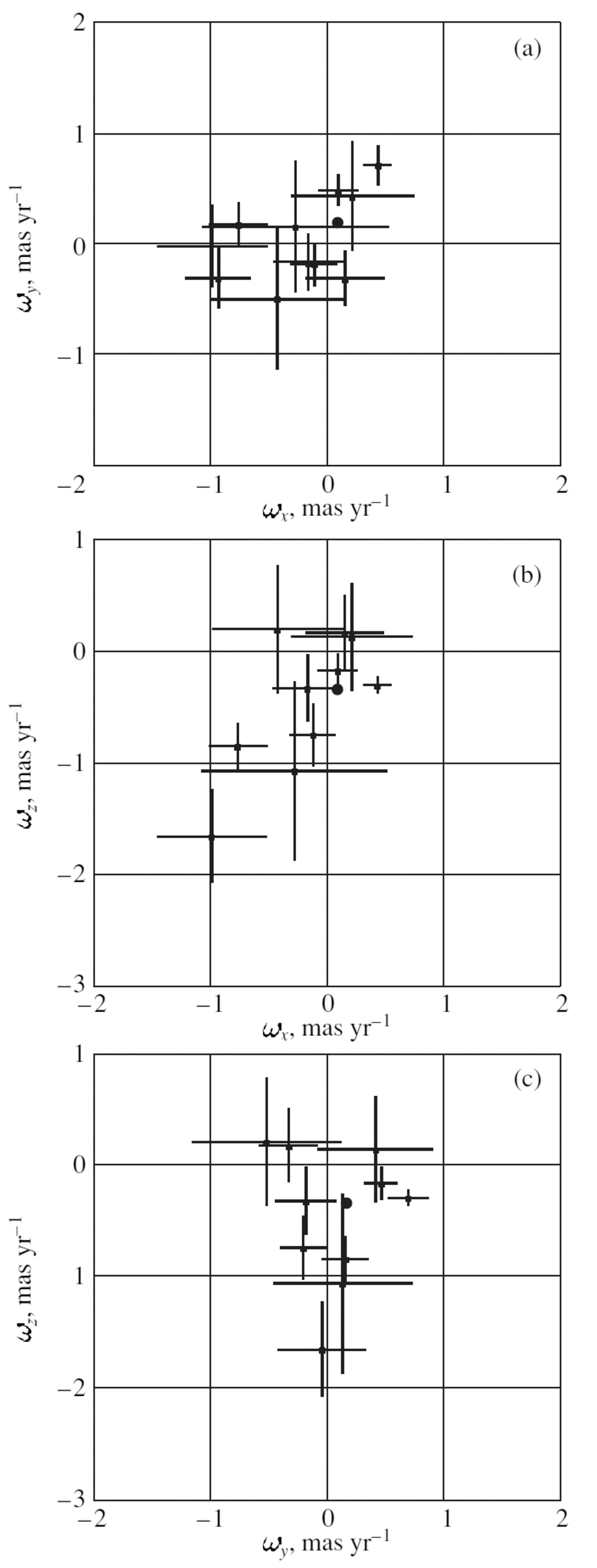}
 \caption{Projections of the individual solutions (without HST) for
the vector ${\hbox {\boldmath $\omega$}}$: (a) onto the $xy$
plane; (b) onto the $xz$ plane, and (c) onto the $yz$ plane. The
filled circle indicates the components of our Mean 1 solution.}
 \label{f1} \end{center} } \end{figure}

Since Ma et al. (1998) adopted the correction
 $\Delta p_1 = -2.84\pm0.04$ mas yr$^{-1}$ to the IAU (1976)
constant of lunisolar precession in longitude, it may be assumed
that $\Delta p_1$ was underestimated. Since the ICRS was
constructed with precisely this value, solution~(6) yields an
``addition'' to this correction. In this case, the correction to
the IAU (1976) constant of lunisolar precession in longitude is
$$
 \displaylines{\hfill \Delta p_1=-3.22\pm0.11~{\hbox {mas yr$^{-1}$}}.\hfill\llap(7)}
$$
Solution (7) is in satisfactory agreement with the most recent
results from the laser ranging of the Moon,
 $\Delta p_1=-3.02\pm0.03$ mas yr$^{-1}$ (Chapront et al. 2002),
and with the results of an analysis of radio interferometric
observations,
 $\Delta p_1=-3.011\pm0.003$ mas yr$^{-1}$ (Fukushima 2003).
On the other hand, if we assume that the lunisolar precession has
no effect on the construction of ICRS and ICRF, then $\Delta
E\not=0$. In this case, we find from Eq.~(5) that
$$
 \displaylines{\hfill \Delta E=+0.38\pm0.10~{\hbox {mas yr$^{-1}$}}.\hfill\llap(8)}
$$
The residual rotation of the Hipparcos catalog that we found is
small. An analysis of the radio-interferometric observations
(Boboltz et al. 2003) performed 9.69 years after (epoch 2000.94)
the construction of ICRS (epoch 1991.25) revealed no significant
effect in the coordinates of radio stars. However, the rotation
components obtained by Boboltz et al. (2003) from the coordinate
differences for 18 radio stars (Kovalevsky et al. 1997) are of
considerable interest:
$$
\displaylines{\hfill
 \varepsilon_{0x}=+0.2\pm2.9~{\hbox {mas}},\hfill\llap(9)\cr\hfill
 \varepsilon_{0y}=+1.9\pm3.2~{\hbox {mas}},\hfill\llap(10)\cr\hfill
 \varepsilon_{0z}=-2.3\pm2.8~{\hbox {mas}}.\hfill\llap(11)\cr
 }
$$
We changed the signs of the components of the vector $\varepsilon$
inferred by Boboltz et al. (2003) in order to have Hipparcos
catalog differences comparable with those analyzed. To compare
results (9)--(11), for example, with the Mean 1 solution, the
quantities $\varepsilon_{0x,y,z}$ must be divided by the epoch
difference. Thus, the difference $\varepsilon_{0z}$, which is
determined with the smallest random error, is largest among
quantities (9)--(11), thereby confirming our $\omega_z$ value. In
general, quantities (9)--(11) agree well with the Mean 1 solution.

Our analysis of the available individual sources for controlling
the ICRS inertiality shows that they are of little use for
analyzing future projects, such as GAIA, SIM, etc., in which a
microacrsecond accuracy is expected to be reached (Kovalevsky et
al. 1999). The idea that the inertiality must be controlled using
quasar observations from a spacecraft directly during its flight
(Metz and Geffert 2004) seems most promising.

\bigskip
\leftline {\hskip6mm CONCLUSIONS}
\bigskip
We have confirmed that the error in referring the ICRS to the
inertial reference frame is very small and does not exceed
 $\pm0.25$ mas yr$^{-1}$ (along the three axes).

We showed that the equatorial component
 $\omega_z = -0.35\pm0.09$ mas yr$^{-1}$
of the vector of residual rotation of the ICRS with respect to the
inertial reference frame differs significantly from zero. This
confirms the result of a kinematic analysis of the proper motions
for stars of the ICRS catalogs (Bobylev 2004).

 \medskip
 \leftline {\hskip6mm ACKNOWLEDGMENTS}
 \medskip
This work was supported by the Russian Foundation for Basic
Research (project no. 02--02--16570).

 \bigskip
 \leftline {\hskip6mm REFERENCES}
 \bigskip
 {\small

1. D.A. Boboltz, A.L. Fey, K.J. Johnston, et al., Astron. J. {\bf
126}, 484 (2003).

2. V.V. Bobylev, Pis’ma Astron. Zh. {\bf 30}, 289 (2004) [Astron.
Lett. {\bf 30}, 251 (2004)].

3. V.V. Bobylev, N. M. Bronnikova, and N. A. Shakht, Pis’ma
Astron. Zh. {\bf 30}, 519 (2004).

4. J. Chapront,   M.~Chapront-Touz\'e, and G. Francou, Astron.
Astrophys. {\bf 387}, 700 (2002).

5. A.N. Deutch, Trans. IAU 8, 789 (1954).

6. A.N. Deutch, V.V. Lavdovskii, and N.V. Fatchikhin, Izv. Gos.
Astron. Obs. {\bf 154}, 14 (1955).

7. T. Fukushima, Astron. J. {\bf 126}, 494 (2003).

8. M. Geffert, A. R. Klemola, M. Hiesgen, et al., Astron.
Astrophys. {\bf 124}, 157 (1997).

9. R.B. Hanson, A.R. Klemola, B.F. Jones, et al., Lick Northern
Proper Motion Program: NPM2 Catalog,
http://www.ucolick.org/?npm/NPM2/ (2003).

10. The HIPPARCOS and Tycho Catalogues, ESA SP- 1200 (ESA, 1997).

11. S. Hirte, E. Schilbach, and R.-D. Scholz, Astron. Astrophys.,
Suppl. Ser. 126, 31 (1996).

12. V.S. Kislyuk, S.P. Rybka, A.I. Yatsenko, et al., Astron.
Astrophys. {\bf 321}, 660 (1997).

13. A.R. Klemola, R.B.Hanson, and B.F. Jones, Galactic and Solar
System Optical Astrometry, Ed. by L.V. Morrison and G.F. Gilmore
(Cambridge Univ. Press, Cambridge, 1994), p. 20.

14. A.R. Klemola, B.F. Jones, and R.B.Hanson, Astron. J. {\bf 94},
501 (1987).

15. J. Kovalevsky, {\it JOURNE\'ES 1999}, Ed. by M. Soffel and N.
Capitaine (Obs. de Paris, Paris, 1999), p. 103.

16. J. Kovalevsky, L. Lindegren, M.A.C. Perryman, et al., Astron.
Astrophys. {\bf 323}, 620 (1997).

17. J.-F. Lestrade, D.L. Jones, R.A. Preston, et al., Astron.
Astrophys. {\bf 304}, 182 (1995).

18. J.-F. Lestrade, R. A. Preston, D.I. Jones, et al., Astron.
Astrophys. {\bf 344}, 1014 (1999).

19. L. Lindegren and J. Kovalevsky, Astron. Astrophys. {\bf 304},
189 (1995).

20. C. Ma, E.F. Arias, T.M. Eubanks, et al., Astron. J. {\bf 116},
516 (1998).

21. M. Metz and M. Geffert, Astron. Astrophys. {\bf 413}, 771
(2004).

22. I. Platais, T.M. Girard, V. Kozhurina-Platais, et al., Astron.
J. {\bf 116}, 2556 (1998a).

23. I. Platais, T.M. Girard, W.F. van Altena, et al., Astron.
Astrophys. {\bf 304}, 141 (1995).

24. I. Platais, V. Kozhurina-Platais, T.M. Girard, et al., Astron.
Astrophys. {\bf 331}, 1119 (1998b).

25. S.P. Rybka and A.I. Yatsenko, Kinemat. Fiz. Neb. Tel 13, 70
(1997a).

26. S.P. Rybka and A.I. Yatsenko, Astron. Astrophys., Suppl. Ser.
{\bf 121}, 243 (1997b).

27. H.-J. Tucholke, P. Brosche, and M. Odenkirchen, Astron.
Astrophys., Suppl. Ser. {\bf 124}, 157 (1997).

28. J. Vondr\'ak, C. Ron, and I.~Pe\v sek, Astron. Astrophys. {\bf
319}, 1020 (1997).

29. Zi Zhu, Publ. Astron. Soc. Jpn. {\bf 53}, L33 (2001).

30. Zi Zhu, {\it JOURNE\'ES 2003}, Ed. N.~Capitaine (Obs. de
Paris, Paris, 2003).
 }

\end{document}